\pdfoutput=1
\documentclass[aip,jcp,reprint,floatfix,a4paper]{revtex4-1}  
\usepackage{cmap}
\usepackage[T1]{fontenc}
\usepackage[Euler]{upgreek}
\usepackage{times}
\usepackage{helvet}
\usepackage{fixmath}
\DeclareSymbolFont{UPM}{U}{eur}{m}{n}
\DeclareMathSymbol{\partial}{0}{UPM}{"40}
\usepackage{graphicx}
\usepackage{amsmath,esint}
\usepackage{geometry}
\usepackage[stretch=15,shrink=15,step=1]{microtype}
\SetProtrusion{encoding={*},family={*},series={*},size={6,7}}
              {1={ ,550},2={ ,300},3={ ,300},4={ ,300},5={ ,300},
               6={ ,300},7={ ,400},8={ ,300},9={ ,300},0={ ,300}}
\usepackage[nodayofweek]{datetime}
\newdateformat{myDate}{\THEDAY\ \monthname[\THEMONTH] \THEYEAR}
\usepackage{bm}
\usepackage{placeins}
\usepackage{csquotes}
\usepackage{mathtools}
\usepackage{bbm}
\DeclareMathOperator{\sgn}{sgn}
\usepackage{hyperref}
\hypersetup{colorlinks=true,citecolor=black, filecolor=black,linkcolor=black,urlcolor=black}

\begin{document}
\preprint{}
\title[]{{\color{black}{Numerical Evidence for Thermally Induced Monopoles}}}

\author{P. Wirnsberger}
\affiliation{
{\footnotesize{Department of Chemistry, University of Cambridge, Cambridge CB2 1EW,
United Kingdom.}}%
}%

\author{D. Fijan}
\affiliation{
{\footnotesize{Department of Chemistry, University of Cambridge, Cambridge CB2 1EW,
United Kingdom.}}}%
\affiliation{
{\footnotesize{Current address: Department of Chemistry, University of Oxford, Oxford OX1 3QZ,
United Kingdom.}} \\
}%

\author{R.~A. Lightwood}
\affiliation{
{\footnotesize{Department of Chemistry, University of Cambridge, Cambridge CB2 1EW,
United Kingdom.}}}%

\author{A. \v{S}ari\'c}%
\affiliation{
{\footnotesize{Department of Chemistry, University of Cambridge, Cambridge CB2 1EW,
United Kingdom.}}}%
\affiliation{
{\footnotesize{Current address: Department of Physics and Astronomy, Institute for
the Physics of Living Systems, University College London, WC1E 6BT, United
Kingdom.}} \\%
}%

\author{C. Dellago}
\affiliation{%
{\footnotesize{Faculty of Physics, University of Vienna, 1090 Vienna, Austria.}}
\\
}%

\author{D. Frenkel}%
\email[Corresponding author: ]{df246@cam.ac.uk}
\affiliation{
{\footnotesize{Department of Chemistry, University of Cambridge, Cambridge CB2
1EW, United Kingdom.}}%
}%

\date{\today}
 
\maketitle

{\bf 
Electrical charges are conserved. The same would be expected to hold for 
magnetic charges, yet magnetic monopoles have never been observed.
It is therefore surprising that the laws of non-equilibrium thermodynamics, 
combined with Maxwell's equations, suggest that colloidal particles heated or cooled 
in certain polar or paramagnetic solvents may behave as if they carry an 
electrical/magnetic charge~\cite{Frenkel2016}. Here we present numerical simulations 
that show that the field distribution around a pair of such heated/cooled 
colloidal particles agrees quantitatively with the theoretical predictions 
for a pair of oppositely charged electrical or magnetic monopoles. However, in other respects, 
the non-equilibrium colloids do not behave as monopoles: they cannot be moved by a homogeneous applied field.  
The numerical evidence for the monopole-like fields around heated/cooled colloids is crucial because the experimental and 
numerical determination of forces between such colloids would be complicated by the 
presence of other effects, such as thermophoresis.}

The existence of quasi-monopoles in a system of heated or cooled colloids in a polar or paramagnetic 
fluid follows directly from non-equilibrium 
thermodynamics, combined with the equations of electro/magneto-statics~\cite{Frenkel2016}.
Although suggested theoretically, they have thus far not been studied experimentally. 
The present paper provides numerical evidence  indicating that the predicted effects are real and robust. 
In what follows, we consider the case of thermally induced quasi-monopoles in a
dipolar liquid, but all our results also apply to paramagnetic liquids.
It has been shown that a thermal gradient will create an electrical field in a liquid of 
dipolar molecules with sufficiently low symmetry~\cite{Bresme2008, Romer2012}. 
In the absence of any external electric field, a heated or cooled colloid placed in such a liquid, 
will generate the field~\cite{Bresme2008, Armstrong2015a, Lee2016}
\begin{equation}
\label{eq:ETP}
\mathbold E_\text{TP}(\mathbold r) = S_\text{TP} \nabla T(\mathbold r),
\end{equation}
where $T(\mathbold r)$ is the temperature and $S_\text{TP}$ the
thermo-polarisation coefficient, with a magnitude that is not known \textit{a priori}. 
For water near room temperature, $S_\text{TP}$ has been estimated to 
be $S_\text{TP} \approx 0.1~\text{mV/K}$~\cite{Armstrong2015a,Wirnsberger2016}.

Let us next consider the electrical polarisation around a heated (or cooled) colloidal particle. 
In steady state the temperature profile at a distance $r$ from the centre of an isolated, spherical colloid of radius $R$ satisfies
\begin{equation}
\label{eq:Tr}
T(r)= T_\infty +(T_R-T_\infty){\frac{R}{r}},
\end{equation}
and hence
\begin{equation}
\label{eq:Er}
{\mathbold E}_\text{TP}(\mathbold r) = -S_\text{TP}(T_R-T_\infty){\frac{R}{r^2}} {\hat {\mathbold r}},
\end{equation}
where $T_\infty$ is the temperature in the bulk liquid and ${\hat {\mathbold r}}$ the radially outward pointing unit vector. 
Note that $E_\text{TP}$ decays as $1/r^2$. Using Gauss's theorem, we can then write
\begin{equation}
\label{eq:qsphere}
\oiint \mathbold E_\text{TP}(\mathbold r) \cdot \mathrm d\mathbold S = -4\pi S_\text{TP} (T_R-T_\infty) R \equiv {q_\text{TP}\over\epsilon_0},
\end{equation}
where $\epsilon_0$ is the dielectric permittivity of vacuum. 
In words: the flux through a closed surface around a neutral colloid is non-zero, 
and is equal to the flux due to an apparent charge $q_\text{TP} = -4\pi \epsilon_0 S_\text{TP} (T_R-T_\infty)R$. 
Note that the effective charge is proportional to the radius of the particle, hence larger colloids will have a larger apparent charge. 

\begin{figure}[b!]
  \centering
  \includegraphics{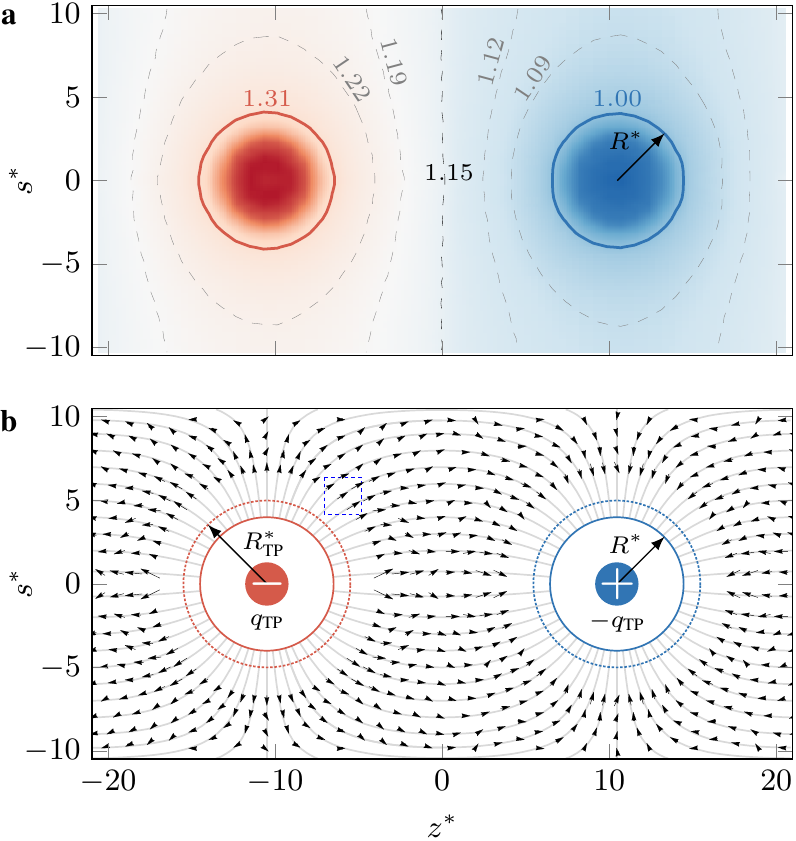}
  \caption{ 
            \textbf{a} Cylindrically averaged temperature profile with symmetry axis $z^*$, 
                       perpendicular direction $s^*$, and isosurfaces (solid and dashed lines) around 
                       two colloids of radius $R^*$, one heated and the other one cooled.
            \textbf{b} Cylindrically averaged field lines generated by two point charges, $\pm q_\text{TP}$,
                       with periodic boundary conditions. The superimposed arrows indicate the average dipolar 
                       orientations obtained from the simulations. Averages were calculated inside small volumes (dashed rectangle). 
                       To avoid spurious boundary effects, we did not consider dipoles within a radius $R^*_\text{TP}$ from the 
                       center of either colloid.}
\label{fig:TMur}
\end{figure}

To verify the existence of thermally induced charges numerically, we performed non-equilibrium molecular dynamics (NEMD)
 and equilibrium MD simulations of a heated and a
cooled colloid immersed into a modified (`off-centre') Stockmayer fluid~\cite{Stockmayer1941}, 
consisting of particles with a point dipole and a Lennard-Jones (LJ) centre displaced along the direction 
of the dipole moment (Appendix~\ref{app:model}).
This displacement is controlled by a parameter $\alpha$. A non-zero value of $\alpha$  is necessary 
to observe the effect~\cite{Romer2012}, as
molecules would otherwise have no preferred thermo-molecular orientation.
An important property of our model fluid is that $S_\text{TP}$ is effectively constant in the temperature and density
range investigated (Appendix~\ref{app:stp}), thereby facilitating the analysis as compared to 
the polar models considered previously~\cite{Bresme2008, Muscatello2011a,Armstrong2013, Daub2014,Daub2016, Armstrong2015a, Iriarte-Carretero2016, Wirnsberger2016}.
The temperature gradient is sustained by continuously pumping energy into the hot colloid and removing it from 
the cold one such that the overall system energy is
constant~\cite{Wirnsberger2015}.

In our numerical simulations, we chose a geometry in which the two colloids  
are located on the $z$-axis in a system with periodic boundary conditions. 
As a first test of the theory, we measured two-dimensional steady state profiles for the
temperature and the average dipolar orientations, both shown in
Fig.~\ref{fig:TMur}. Quantities labeled with an asterisk are expressed in reduced units, defined in the {\em Methods} section.
To improve statistics, we computed cylindrical averages (indexed by $z$ and $s$), although the underlying problem does not exhibit full
radial symmetry in the $xy$--plane due to effects of periodic boundary conditions. 
However, as the theoretical predictions were also cylindrically averaged, the 
comparison between simulation and theory is still valid.
The dashed vertical line going through the origin of Fig.~\ref{fig:TMur}a corresponds to the equilibrium (or bulk) temperature $T_\infty$. 
With the temperature values of the specific contour lines shown in the figure and a value
of $S^*_\text{TP}  = (0.216 \pm 0.022)$ computed in the vicinity of the origin (Appendix~\ref{app:stp}), we can employ
equation~(\ref{eq:qsphere}) to obtain an estimate  
of $q^*_\text{TP} \approx -0.14$ for the thermally induced charge.
If we use the LJ parameters of SPC/E water~\cite{Berendsen1987} for the unit conversion, this
corresponds to $q_\text{TP}  \approx 5.4  \times 10^{-3} q_e$, where $q_e$ is the charge of an electron.

Figure~\ref{fig:TMur}b shows the average dipolar orientations
superimposed onto the field lines generated by two virtual point charges located at the
centres of the colloids. 
To single out the thermally induced alignment 
from contributions already present in equilibrium, e.g. the alignment caused by
surface layering of solvent molecules in the vicinity of the colloids, we
measured equilibrium orientations in a separate simulation and subtracted them from the non-equilibrium result.
This procedure assumes that the coupling between the various contributions to the total field is negligible.
We found this assumption to be reasonable everywhere apart from the immediate vicinity of the colloids. Therefore we excluded 
the first layer of solvent molecules, i.e. all particles within a distance of $R_\text{TP}^*= 5$ from the colloid centres, from the averaging.
The precise value of $R_\text{TP}$ does not matter as long as it is chosen sufficiently large. We picked the smallest
value that allows us to single out the effect. As we can see, the  dipoles are aligned very well with 
the field lines generated by two point charges in a periodic system. 

As a more quantitative test of the theory, we measured the electric field induced by the temperature gradient.
To improve the statistical accuracy of our results, we average the field over planes
perpendicular to the symmetry axis, such that all contributions apart from $E_{z,\text{TP}}$ cancel out.
The system behaves as if  the two charges of opposite sign are distributed over thin spherical shells of 
radius $R_\text{TP}$, as depicted in Figs~\ref{fig:Ez}a and b. For this geometry,  we obtain the analytical solution 
for the electrical field (see Appendix~\ref{app:model}):
\begin{equation}
\label{eq:EzTPfull}
\frac{\langle E_{z,\text{TP}}(z) \rangle}{\tilde E}  = 
 \begin{cases} 
   -1                                  &\mbox{if $|z| > z_\text{c}+R_\text{TP}$,} \\
   +1                                  &\mbox{if $|z| < z_\text{c}-R_\text{TP}$,} \\
   (z-z_\text{h})/R_\text{TP}          &\mbox{if $|z-z_\text{h}| \leq R_\text{TP}$, } \\
   (z_\text{c} - z)/R_\text{TP}        &\mbox{otherwise,} 
  \end{cases}
\end{equation}
where $z_\text{h/c} = \mp L/4$ denote the locations of the hot and cold colloid, respectively, $L$ is the box length in the $z$-direction,
$\tilde E = q_\text{TP}/(2 A\epsilon_0)$ is the constant value of the averaged field between the colloids, and $A = L^2/4$ is the cross-sectional area.
The left-hand side of the above expression can be related to the average dipole density such that~\cite{Wirnsberger2016}
\begin{equation}
\label{eq:EzTSim}
\langle E_{z,\text{TP}}(z) \rangle = -{\langle  {\rho}_{\mu}(z)
-{\bar \rho}_{\mu} \rangle\over\epsilon_0},
\end{equation}
where ${\bar \rho}_{\mu} = 1/L \int \mathrm dz
\langle {\rho}_{\mu}(z) \rangle$ is the box average of $\langle {\rho}_{\mu}(z)
\rangle$. 
Equation~(\ref{eq:EzTSim}) enables us to link the theory and NEMD simulations quantitatively. 
We can estimate the right-hand side of the above equation readily by sampling the instantaneous 
dipole orientations and performing temporal and spatial averaging for slabs perpendicular to the symmetry axis.
Using equation~(\ref{eq:EzTPfull}), we can then infer the value of ${\tilde E}$ from our results and 
obtain an independent numerical estimate of $q_\text{TP}$, in addition to the one provided 
by equation~(\ref{eq:qsphere}). Observing a good agreement for both estimates would provide strong 
support for the theory, since it would suggest that Gauss's theorem can be applied to arbitrary 
volumes enclosing the colloids, just as if they carried real Coulomb charges. We note, however, 
that there is an important conceptual difference between estimating the charge using equation~(\ref{eq:EzTPfull}) versus
equation~(\ref{eq:qsphere}): the latter already \textit{assumes} that
equation~(\ref{eq:ETP}) holds whereas the former \textit{validates}
it. 

\begin{figure}
 \centering
  \includegraphics{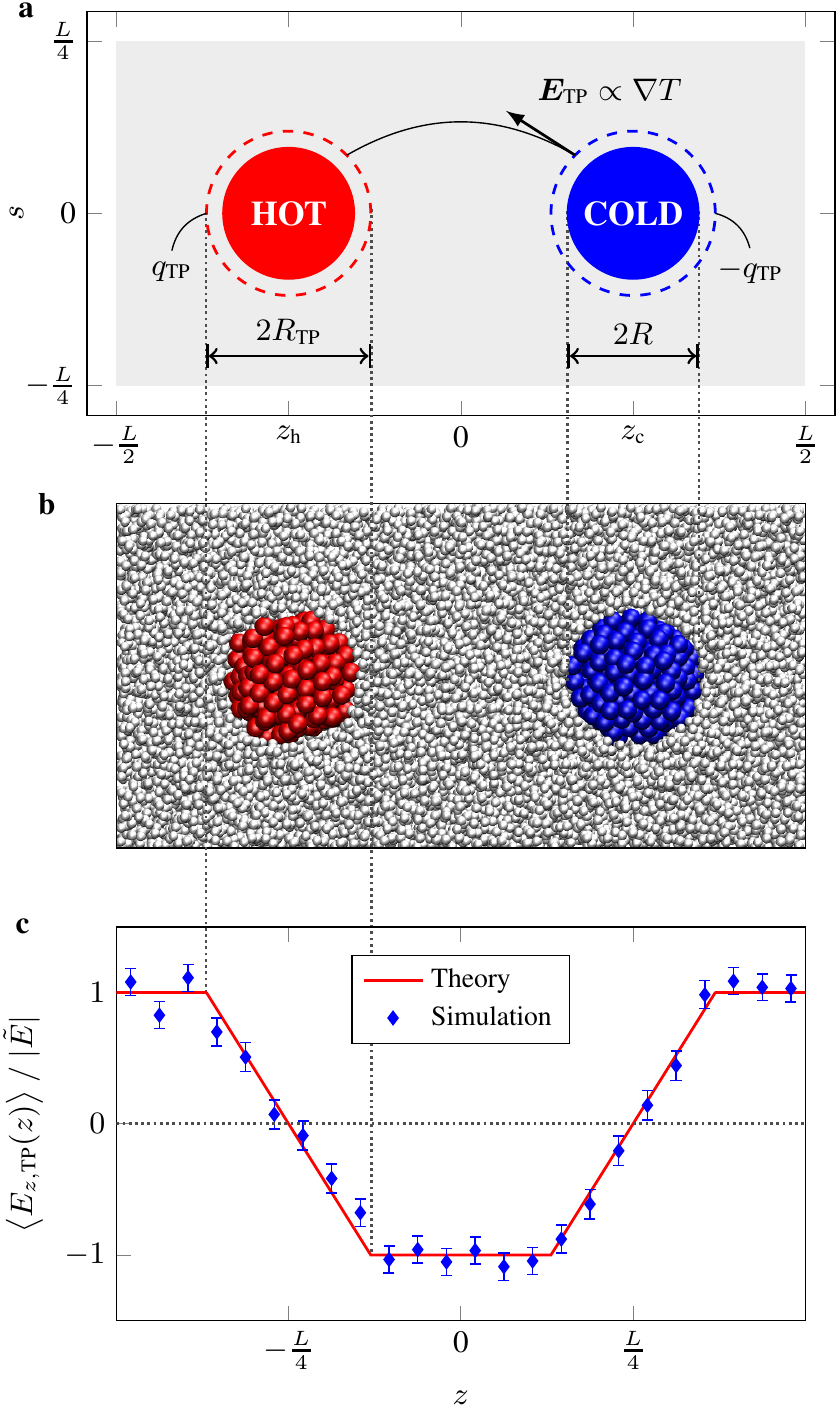}
  \caption{
  \textbf{a}     Illustration of the setup. 
                 The dashed lines of radius $R_\text{TP}$ enclosing the hot and cold colloids represent 
  		 infinitesimally thin spherical shells carrying the induced charges $\pm q_\text{TP}$. 
		 The black solid line illustrates a field line and the arrow represents a field vector.
  \textbf{b}     A typical configuration obtained from simulation showing the colloids immersed into the solvent particles. 
  \textbf{c}     Thermally induced field averaged over slabs perpendicular to the symmetry axis. 
		 The simulation results (blue symbols) were calculated from the averaged dipole density excluding 
                 two balls of radius $R_\text{TP}$ centred around the colloids. 
                 The solid line shows the theoretical prediction given 
		 by equation~(\ref{eq:EzTPfull}). The dotted vertical and horizontal lines
		 were added to guide the eye and to highlight the symmetry of the induced field.
   }
\label{fig:Ez}
\end{figure}

Figure~\ref{fig:Ez}c shows the steady state result for the
spatial variation of the averaged field calculated
according to equation~(\ref{eq:EzTSim}).
Equilibrium averages were subtracted and solvent particles within a distance of $R_\text{TP}$ from the colloid 
centres excluded from the averaging, which makes the effective radius of the charge distribution essentially 
an input parameter of our model. We can see that the simulation data are in excellent 
agreement with the theoretical expression~(\ref{eq:EzTPfull}):
the average field is constant in the fluid region and changes linearly within a
distance of $R_\text{TP}$ from the colloid centre. From the plateau in the centre we estimate 
$\tilde E^* = (-1.96 \pm  0.20)\times 10^{-3}$ for the regions where the field is constant, 
and find a value of $q_\text{TP} = (5.27\pm0.54) \times 10^{-3} q_e$ for the thermally induced charge.
Both estimates for $q_\text{TP}$ are in excellent agreement.
The sign of $q_\text{TP}$ can be controlled either by changing the rate of
energy, $\mathcal {F}$, supplied to or withdrawn from the colloid (flipping hot
and cold) or by changing $\alpha$, such that $\sgn(q_\text{TP}) = \sgn (\alpha) \sgn (\mathcal F)$. 

A key question is whether the effective electrical or magnetic charge of colloidal monopoles can be measured in experiments. 
The present simulations suggest that, at the very least the effect of the monopole fields on probe charges (or dipoles) should be observable. 
Of course, it would be attractive to make the effect as large as possible by increasing the temperature difference between the particle and the solvent. 
However, the temperature range is limited by the fact that extreme heating or cooling will bring the system out of the linear-response regime - 
and possibly even induce phase transitions in the solvent. Moreover, the colloidal monopoles differ in an important respect from true monopoles: 
they cannot be moved by a uniform external field~\cite{Frenkel2016}. It is therefore  tempting (be it slightly frivolous) to call such 
colloidal monopoles `quacks', as they quack like a duck (i.e. they create a field similar to that of a real monopole), 
but they don't swim like a duck (they cannot be used to transport charge). One of the main effects that may obscure observation 
of the Coulomb-like interaction between oppositely heated colloids is thermophoresis, which will also cause colloids to move in the 
temperature gradient caused by another colloid. However, at least in the linear regime, this effect should cause otherwise identical 
but oppositely heated colloids to move {\em in the same direction} with respect to the fluid rather than with respect to one another. 
Finally, there are many open questions about the practical consequences of the existence of thermal monopoles. 
It is, for instance, conceivable that such particles in an electrolyte solution will get `decorated' with real charges, 
and thereby acquire real charge (opposite and equal to the `thermal' charge) that can be dragged along. 
That charge should respond to a uniform external field: the resulting electro-osmotic flow would cause motion of the colloids.

\section*{\textbf{Acknowledgements}}
{\small
P.~W. acknowledges the Austrian Academy of Sciences for financial support through a DOC Fellowship,
and Martin Neumann, Chao Zhang, Michiel Sprik, Aleks Reinhardt, Carl P\"olking and Tine Curk
for many invaluable discussions.
We further acknowledge financial support of the Austrian Sciences Fund FWF 
within the SFB Vicom [Project No. F41] (C.~D.) and the EU ITN Project NANOTRANS [Grant No. 674979] (D.~F.).
The results presented here have been
achieved in part using the Vienna Scientific Cluster (VSC). 
}

\section*{\textbf{Author contributions}}
{\small
The research was planned by P.~W., C.~D. and D.~F. based on a theoretical suggestion by D.~F.
Simulations were carried out by P.~W. with assistance of R.~A.~L., D.~Fijan and A.~\v{S}.
The manuscript was written by P.~W., A.~\v{S}., C.~D. and D.~F.}

\section*{\textbf{Methods}}
{\small \textbf{Simulation setup.}
All equilibrium and non-equilibrium MD simulations were performed using the
software package LAMMPS~\cite{Plimpton1995} (version 14Jun16). We employed a
fully periodic rectangular simulation box with dimensions $(L_x,L_y,L_z) = (L/2,L/2,L)$, where $L = 41.93\sigma$, 
containing $13422$ solvent particles of LJ diameter $\sigma$, which defines 
the unit of length, and two colloids centred at $\mathbold r_\text{h/c} = (0,0,\mp L/4)$.
Each colloid was modelled with an elastic network of 201 beads with $2808$ harmonic
springs connecting nearest, second-nearest and third-nearest neighbours. The
initial colloid configuration was cut out of an fcc lattice with a density of
$0.75/\sigma^3$ matching the solvent density. Springs were then added to
all beads within a distance of $R = 4\sigma$ of the centre of mass positions of the
two colloids. The equilibrium distances of the harmonic spring potentials were taken to 
be the initial bead separations and the
spring constant was set to $5 \epsilon/\sigma^2$, where the LJ parameter $\epsilon$ 
defines the energy scale. During the simulation the colloids
were held in place by two additional stiff harmonic springs ($100 \epsilon/\sigma^2$) tethering the
centres of mass to the equilibrium positions.
The solvent molecules were modelled as modified Stockmayer particles 
consisting of a point dipole, located at the particle's centre of mass, and
a shifted LJ centre. We displaced
the LJ centre from the dipole by $\upDelta \mathbold r = \alpha
\hat {\mathbold \mu}$, where $\alpha = -\sigma/4$ controls the
asymmetry and $\hat{\mathbold \mu}$ is the unit vector of the dipole moment
$\mathbold \mu$. 
This modification leads to additional torque
contributions which are summarised in Appendix~\ref{app:modstock}.
We used the relations $\mathbold \mu^* = \mathbold \mu / \sqrt{4 \pi \epsilon_0 \epsilon \sigma}$ 
and $q^* = q / \sqrt{4 \pi \epsilon_0 \epsilon \sigma}$ to non-dimensionalise 
dipole moment and charge and set both $4\pi\epsilon_0$ and $\mu^*$ to unity. 
The colloidal bead-solvent interactions were modelled with a 
LJ potential using the same parameters, $\epsilon$ and $\sigma$, as for
the solvent-solvent interactions, and both solvent particles and colloidal beads have the same mass $m$.
Electrostatic interactions were treated with Ewald summation and tin-foil boundary
conditions~\cite{DeLeeuw1980}. Cutoff radii for all LJ and real space Coulomb interactions 
were set to $8\sigma$ and the $k$-space settings were chosen such that
the relative accuracy of the force was approximately $10^{-5}$, as estimated 
with the formulas provided in ref.~\onlinecite{Wang2001}. The equations of
motion were integrated using a timestep of $\upDelta t = 0.002
\tau $, where $\tau = \sigma \sqrt{ m/\epsilon}$ is the unit of time.

\textbf{Equilibration.} 
The initial lattice structure was
equilibrated in the \textit{NVT} ensemble for a period of $2 \times 10^3 \tau$ using a
Nos\'e--Hoover thermostat~\cite{Nose1984, Hoover1985} with a relaxation time 
of $0.5\tau$ and a target temperature of $T_\infty = 1.15 \epsilon/k_\text{B}$, where
$k_\text{B}$ is the Boltzmann constant which was set to unity.
Subsequently, all particle velocities of the last configuration were 
rescaled to match the average kinetic energy of the \textit{NVT} run, 
which was followed by a $2 \times 10^3\tau$ long \textit{NVE} equilibration run. 
A heat flux was then imposed onto the sytsem using the
eHEX/a algorithm~\cite{Wirnsberger2015}, where the rate of energy supplied
to the hot (and withdrawn from the cold) colloid was set to $\mathcal{F} = 52.75
\epsilon/\tau$. After waiting for a period of $10^4 \tau$ for any transient
behaviour to disappear and the system to reach a steady state, we started the
$1.5 \times 10^5 \tau$ long production run and stored snapshots of the trajectory for
further post-processing of translational, kinetic temperature and dipole orientations.
In addition, we carried out a $1.5 \times 10^5 \tau$ long \textit{NVE} simulation in order 
to subtract non-vanishing equilibrium averages of the spatially averaged field 
and the dipolar orientations from the NEMD results. The relative increase in the
total energy throughout the entire NEMD production run (75 million timesteps) was approximately $0.14\%$, 
which is comparable to the value of $0.12\%$ for the equilibrium production run.

\textbf{Statistical accuracy.}
The size of each error bar in Fig.~\ref{fig:Ez}c represents twice the standard deviation of the
mean value which was calculated as the difference between the non-equilibrium and the equilibrium averages.
For the individual production run we computed field averages according to the
following protocol:
at regular time intervals of $\delta t = 50 \upDelta t$
we computed $\langle E_{z,\text{TP}}(z) \rangle$ according to equation~(\ref{eq:EzTSim}),
excluding dipoles within a distance of $R_\text{TP}$ from the colloid centres.
We then averaged $\langle E_{z,\text{TP}}(z) \rangle$ over slabs of width $\upDelta z = L/24$
which are centred around the points $z_i = -L/2 + (i-1/2)\upDelta z$, where $i=1,\ldots,24$.
The resulting instantaneous spatial averages are denoted by $E_i^{m}$, where $m = 1,\ldots, M$ 
indexes the simulation time according to $t^m = m \delta t$
and $M = 1.5\times10^6$ is the total number of configurations considered.
From the resulting time series $\{E_i^1,\ldots,E_i^M\}$ we computed the mean value, $\bar E_i$, 
for each bin and estimated its standard deviation, $\bar {\sigma}_i$,
using block average analysis.
Errors for the final results ${\bar E}_{i,\text{TP}}  = {\bar E}_{i,\text{NEMD}} - {\bar E}_{i,\textit{NVE}}$ shown in the plot
were calculated as the square root of the total variance ${\bar \sigma}^2_{i,\text{NEMD}} + {\bar \sigma}^2_{i,\textit{NVE}}$,
assuming that the production runs were statistically independent.
The quantity $\tilde E$, appearing in equation~(\ref{eq:EzTPfull}), was computed from the 
slabs with index $j \in \{1,2,11,12,13,14,23,24\}$ using the relation $\tilde E = -1/8 \sum_j |{\bar E}_{j,\text{TP}}|$.
These slabs correspond to the region of constant average field between the colloids. 
Errors were propagated assuming that the terms in the sum are statistically independent such that 
the error $\sigma_{\tilde E}$ is given by the square root of
$1/8  \sum_j {\bar \sigma}^2_{j,\text{TP}}$. The estimate of
$q_\text{TP}$ follows from multiplication of $\tilde E$ by 
the constant factor $2 A \epsilon_0$. 
The error bar for the estimate of $q_\text{TP}$ obtained with equation~(\ref{eq:qsphere})
is omitted since we do not have error estimates for the temperature contour lines shown in Fig.~\ref{fig:TMur}a.
The computation of $S_\text{TP}$ involves additional simulation data and is explained in Appendix~\ref{app:stp}.

\appendix
\numberwithin{equation}{section}
\numberwithin{figure}{section}
\numberwithin{table}{section}
\renewcommand{\thefigure}{\Alph{section}\arabic{figure}}
\renewcommand{\theequation}{\Alph{section}\arabic{equation}}

\section{\label{app:model}Analytical model for the field}
In this section we derive the analytical model proposed in equation~(\ref{eq:EzTPfull}).
To this end, we first show how the spatial average of the three-dimensional field, $\langle E_z(z) \rangle$,
calculated from the full charge density, $\rho(\mathbold r)$, is related to the
one-dimensional field, $E_\text{1D}(z)$, calculated from the spatially averaged charge density, $\rho_\text{1D}(z)$. The subscript $_\text{TP}$
used in the main text is dropped for notational convenience. We consider periodic boundary conditions (PBCs)
and understand that this is implicitly taken into account whenever an expression of the form
$\mathbold r - \tilde{\mathbold r}$ is evaluated. 

For an arbitrary charge distribution, the field can be calculated as
\begin{equation}
\label{eq:E3D}
\mathbold E(\mathbold r) = -\kappa \nabla_{\mathbold r} \int_\Omega \mathrm d^3 \tilde r\ G(\mathbold r - \tilde{\mathbold r}) \rho(\tilde{\mathbold r}),
\end{equation}
where $\kappa = (4\pi \epsilon_0)^{-1}$ with $\epsilon_0$ being the vacuum permittivity, $\nabla_{\mathbold r} = (\partial_x, \partial_y,\partial_z)$ is the gradient in 
Cartesian coordinates, $\Omega$ denotes the orthogonal simulation box of volume $V=L_x L_y L_z$ and $G(\mathbold r- \tilde{\mathbold r})$ is a modified kernel that 
takes into account periodicity~\cite{Wirnsberger2016}. 
Averaging the $z$-component of the field over planes perpendicular to the $z$-axis yields~\cite{Wirnsberger2016, Yeh2011}
\begin{subequations}
\label{appeq:E3Davg}
\begin{alignat}{2}
\langle E_z(z) \rangle &=&&\quad  \frac{1}{L_x L_y} \int\limits_{-\frac{L_x}{2}}^{\frac{L_x}{2}}  \int\limits_{-\frac{L_y}{2}}^{\frac{L_y}{2}} \mathrm dx \mathrm dy\ E_z(\mathbold r) \\
                       &=&& \quad - \kappa 
\frac{\partial}{\partial z} \int\limits_{-\frac{L_z}{2}}^{\frac{L_z}{2}} \mathrm d\tilde z
\underbrace{ \frac{1}{L_x L_y} \int\limits_{-\frac{L_x}{2}}^{\frac{L_x}{2}}  \int\limits_{-\frac{L_y}{2}}^{\frac{L_y}{2}} \mathrm d\tilde x \mathrm d\tilde y \  \rho(\tilde{\mathbold r}) }_{= \ \rho_\text{1D}(\tilde z)}   \nonumber\\
                       & && \times \quad \underbrace{ \int\limits_{-\frac{L_x}{2}}^{\frac{L_x}{2}} \int\limits_{-\frac{L_y}{2}}^{\frac{L_y}{2}} \mathrm dx \mathrm dy \ G(\mathbold r- \tilde{\mathbold r})}_{ =\ G_\text{1D}(z-\tilde z)} \\
                       &=&&\quad -\kappa \int\limits_{-\frac{L_z}{2}}^{\frac{L_z}{2}} \mathrm d \tilde z \ G'_\text{1D}(z-\tilde z) \rho_\text{1D}(\tilde z) \\
		       &=&&\quad \frac{1}{\epsilon_0} \int\limits_{-\frac{L_z}{2}}^{z} \mathrm d\tilde z\ \rho_\text{1D}(\tilde z)  + \frac{1}{\epsilon_0}  \underbrace{\frac{1}{L_z} \int\limits_{-\frac{L_z}{2}}^{\frac{L_z}{2}} \mathrm d\tilde z\ \tilde z \rho_{\text{1D}} (\tilde z)}_{ = P_z} \label{appeq:E3Davgsplit} \\
                       &=&&\quad E_\text{1D}(z),
\end{alignat}
\end{subequations}
where
\begin{equation}
\label{appeq:G1D}
G_\text{1D}(z) = 2 \pi \left[ -|z| + \frac{z^2}{L_z} + \frac{L_z}{6}  \right]
\end{equation}
is the spatially averaged kernel for PBCs
and $P_z$ the $z$-component of the average box dipole density.  

Next, we work out the averaged charge density and compute the field from equation~(\ref{appeq:E3Davgsplit}).
The colloids are modelled by two homogeneously charged, spherical shells of radius $R$ 
(in the main text we refer to this quantity as $R_\text{TP}$). Since all equations involved are linear, we can
decompose the problem and focus on a single colloid. 
If we centre the charge distribution of this colloid around the origin, we can formulate the charge 
density as
\begin{equation}
\label{appeq:RHO3D}
\rho^\text{(1)}(\mathbold r) = \frac{q}{4 \pi R^2} \delta(r-R),
\end{equation}
where $q = \int_\Omega \mathrm d^3 r \rho^\text{(1)}(\mathbold r)$ is the total charge, 
$r$ the distance from the origin and $\delta(r-R)$ the Dirac delta function.
Let us assume that $2 R < L_x = L_y \le L_z$ such that the charge distribution is fully contained  within the reference box.
We then have the freedom to integrate over the largest inscribed cylinder and obtain 
\begin{subequations}
\label{appeq:RHO1D1}
\begin{alignat}{2}
\rho^\text{(1)}_{\text{1D}}(z)   &=&&\quad \frac{1}{L_x L_y} \int\limits_{-\frac{L_x}{2}}^{\frac{L_x}{2}}  \int\limits_{-\frac{L_y}{2}}^{\frac{L_y}{2}} \mathrm dx \mathrm dy\ \rho(\mathbold r) \\
                    &=&&\quad \frac{q}{2 R^2 L_x L_y} \int\limits_{0}^{L_x/2} s \mathrm ds\ \delta\left(  \sqrt{s^2+z^2}-R   \right),
\end{alignat}
\end{subequations}
where $r = \sqrt{ s^2 + z^2}$. Employing a second transformation, $\tau = \sqrt{s^2 +z^2}$ with $s \mathrm ds = \tau \mathrm d\tau$, 
it is straightforward to solve the above integral to find
\begin{equation}
\label{appeq:RHO1D}
\rho^\text{(1)}_{\text{1D}}(z)   =
 \begin{cases}
   \frac{q}{2 R A}  &  \mbox{if $|z| < R$,} \\
   0 &\mbox{otherwise,} 
  \end{cases}
\end{equation}
where $A = L_x L_y$ is the cross-sectional area. The averaged charge density taking into 
account both colloids centred around $z_\text{h}$ and $z_\text{c}$, respectively, is therefore 
given by the piecewise constant function
\begin{equation}
\label{appeq:RHO1DTOT}
\rho_{\text{1D}}(z)   =   \frac{q}{2 R A}
 \begin{cases}
   +1 &  \mbox{if $|z-z_\text{h}| < R$,} \\
   -1 &  \mbox{if $|z-z_\text{c}| < R$,} \\
    0 &\mbox{otherwise.} 
  \end{cases}
\end{equation}
If we plug this result into equation~(\ref{appeq:E3Davgsplit}) and carry out the integration, we obtain
the final result
\begin{equation}
\label{appeq:EzTPfull}
\frac{\langle E_{z}(z) \rangle}{\tilde E}  = 
 \begin{cases} 
   -1                         &\mbox{if $|z| > z_\text{c}+R$,} \\
   +1                         &\mbox{if $|z| < z_\text{c}-R$,} \\
   (z-z_\text{h})/R           &\mbox{if $|z-z_\text{h}| \leq R$, } \\
   (z_\text{c} - z)/R         &\mbox{otherwise,} 
  \end{cases}
\end{equation}
where $\tilde E = q/(2\epsilon_0A)$ is the constant field value for the region between 
the two colloids. 

The quantity $\tilde E$ can be understood easily by applying Gauss's theorem 
to the blue control volume shown in Fig.~\ref{fig:cub}. The charge $q$ in the centre represents the thermally induced charge of the hot colloid. 
Let us denote the surface of this volume by $\partial \Gamma$, the union of the two faces highlighted in 
blue by $\partial \Gamma_\parallel$, and the union of the remaining faces by $\partial \Gamma_\bot$. According to Gauss's theorem
the total charge enclosed by $\partial \Gamma$ is related to the field flux through $\partial \Gamma$ such that
\begin{equation}
\label{appeq:Etilde}
\oiint\limits_{\partial \Gamma} \mathbold E(\mathbold r) \cdot \mathrm d \mathbold S = \frac{q}{\epsilon_0},
\end{equation}
where $\mathrm d\mathbold S$ is the surface normal vector. If we decompose the surface integral 
and recall that the surface normal vector is perpendicular to the field on $\partial \Gamma_\bot$ due to the
periodic setup, we find 
\begin{equation}
\label{appeq:Etildederiv}
\oiint\limits_{\partial \Gamma} \mathbold E(\mathbold r) \cdot \mathrm d \mathbold S =  
\iint\limits_{\partial \Gamma_\parallel} \mathbold E(\mathbold r) \cdot \mathrm d \mathbold S +
\underbrace{\iint\limits_{\partial \Gamma_\bot} \mathbold E(\mathbold r) \cdot \mathrm d \mathbold S}_{ = 0} =
 \langle E_{z,\parallel} \rangle 2 A =  \frac{q}{\epsilon_0}.
\end{equation}
Rearranging terms, we find
\begin{equation}
\label{appeq:Etildefinal}
\tilde E  = \langle E_{z, \parallel} \rangle = \frac{q}{2 \epsilon_0 A},
\end{equation}
which is our final result.

We note that the value of $\tilde E$ is constant and does not 
change if we move the surfaces $\partial \Gamma_\parallel$ along the $z$-axis as long 
as they enclose the charge entirely. Finally, we note that the presence of
the opposite charge $-q$ is already taken into account implicity, which is indicated by 
the multiplication by twice the cross-sectional area $A$ in equation~(\ref{appeq:Etildederiv}). Equivalently, we can think of
the result as the sum of two equal contributions, half from the charge $q$ and the other half from $-q$.
\begin{figure}
 \centering
  \includegraphics[width=7.5cm]{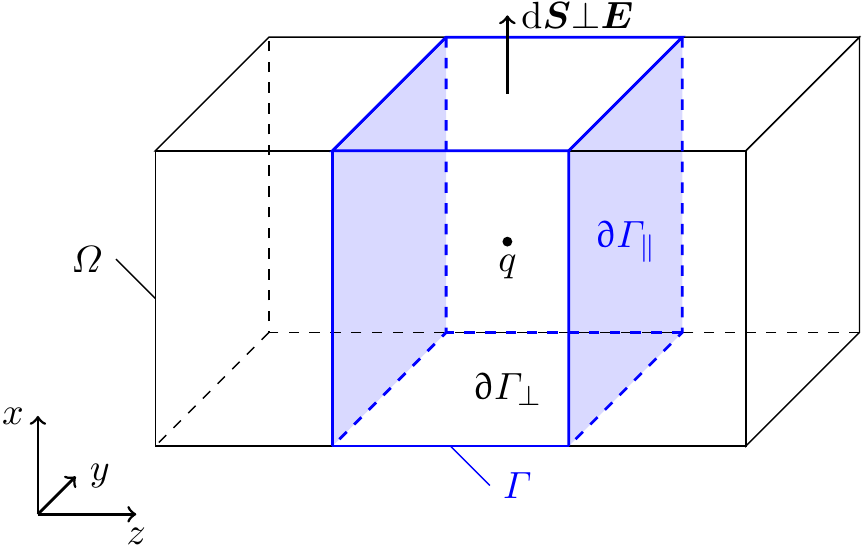}
  \caption{Illustration of the simulation box $\Omega$ containing a charge $q$ 
           located at the centre of a control volume $\Gamma$ (solid and dashed blue lines) to which we apply Gauss's theorem.
           For the two faces of $\Gamma$ denoted by $\partial \Gamma_\parallel$ (highlighted in blue)
           the field $\mathbold E$ has a contribution parallel to the surface normal vector $\mathrm d \mathbold S$,
           whereas it is orthogonal to it on all other faces denoted by $\partial \Gamma_\bot$.
   }
\label{fig:cub}
\end{figure}

\section{\label{app:modstock}Off-centre Stockmayer Model}
Displacing the Lennard-Jones (LJ) centre from the location of the point dipole 
leads to modified forces and torques as compared to the original Stockmayer model~\cite{Stockmayer1941}.
We note that electrostatic contributions are not affected by this modification 
and refer to ref.~\onlinecite{Toukmaji2000} for the relevant expressions. All modifications
of short-ranged interactions related to the perturbation of the LJ centre are governed by a single parameter $\alpha$ and summarised in this section.
\begin{figure}
 \centering
  \includegraphics{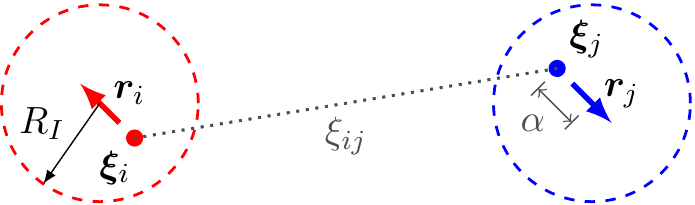}
  \caption{Two solvent particles with dipoles (coloured arrows) located at $\mathbold r_i$ and $\mathbold r_j$, 
           respectively, and displaced LJ centres, $\mathbold \xi_i$ and $\mathbold \xi_j$,
           separated by a distance of $\xi_{ij}$.
           The mass of a solvent particle is distributed homogeneously over a ball of radius $R_I$, as
           illustrated by the dashed circles. The asymmetry 
           in the short-ranged interactions, as compared to the Stockmayer model, is controlled by the parameter $\alpha$.
   }
\label{fig:stockalpha}
\end{figure}

Let us consider the short-ranged, pairwise interactions between two solvent particles as illustrated in Fig.~\ref{fig:stockalpha}.
The point dipoles of mass $m$ are located at the positions $\mathbold r_i$ and $\mathbold r_j$, respectively. 
The mass is distributed homogenously over a ball of radius $R_I = \sigma/2$
such that the moment of inertia is given by $I = 2 m R_I^2/5$, which corresponds to $I^*=0.1$ in reduced units.
The LJ centre is denoted by $\mathbold \xi$ and displaced from the position of the dipole 
by a vector $\upDelta \mathbold r = \mathbold \xi-\mathbold r = \alpha \hat{\mathbold \mu}$, 
where $\hat{\mathbold \mu}$ is the unit vector of the dipole moment $\mathbold \mu$. 
The quantity $\alpha$ allows us to control the level of asymmetry, i.e.
the perturbation to the original Stockmayer model, and we employed a value of $\alpha = -\sigma/4$ 
in all our simulations.

The radially symmetric, pairwise LJ potential is given by
\begin{equation}
\label{appeqq:ulj}
u(r) = 4\epsilon \left[ {\left(  \frac{\sigma}{r} \right)}^{12} -  {\left(  \frac{\sigma}{r} \right)}^{6} \right],
\end{equation}
where $\epsilon$ is the unit of energy. 
For performance reasons, we employed a cutoff of $r_\text{c} = 8\sigma$ for all short-ranged interactions.
The energy contribution for the two particles shown in Fig.~\ref{fig:stockalpha} is
therefore given by $u_{ij} = u(\xi_{ij}) \Uptheta(r_\text{c} - \xi_{ij})$,
where $\Uptheta(r)$ is the Heaviside function and $\mathbold \xi_{ij} = \mathbold \xi_j - \mathbold \xi_i$.
Taking the negative gradient of the energy with respect to $\mathbold \xi_{i}$ and
applying a cutoff, we obtain the force
\begin{equation}
\label{appeq:flj}
\mathbold f_{ij} = \Uptheta(r_\text{c}-\xi_{ij} ) 24 \sigma \epsilon \left[ {\left(  \frac{\sigma}{\xi_{ij}} \right)}^{6} - 2 {\left(  \frac{\sigma}{\xi_{ij}} \right)}^{12} \right]
 \frac{\mathbold \xi_{ij}}{\xi_{ij}^2}
\end{equation}
acting on particle $i$ with the corresponding force $\mathbold f_{ji} = - \mathbold f_{ij}$ acting on particle $j$.
The short-ranged contributions to the torques acting on these particles are then simply given by
\begin{equation}
\label{appeq:tij}
\mathbold \tau_{ij} = \alpha \hat {\mathbold \mu_i} \times \mathbold f_{ij}
\end{equation}
and 
\begin{equation}
\label{appeq:tji}
\mathbold \tau_{ji} = \alpha \hat {\mathbold \mu_j} \times \mathbold f_{ji},
\end{equation}
respectively, where $\times$ denotes the cross product between two vectors. In the limit $\alpha \to 0$
these torque contributions vanish such that we recover the original Stockmayer model.

\section{\label{app:stp}Estimation of {\normalfont $S_\text{TP}$} } 
We estimated the thermo-polarisation coefficient using the relation
\begin{equation}
\label{appeq:stp}
S_\text{TP}(z) =  \frac{  \langle E_{z,\text{TP}}(z)  \rangle  }  {  \partial_z  \big\langle T(z) \big\rangle },
\end{equation}
where $ {  \partial_z  \big\langle T(z) \big\rangle } $ denotes the gradient of the temperature averaged over planes perpendicular to the $z$-axis (see Fig.~\ref{fig:trho1d}).
The simulation data reveals a perfectly linear profile in the vicinity of the origin,
such that $ \beta \equiv  \partial_z  \big\langle T(z) \big\rangle$ is constant. We recall that the field value is $\tilde E$ in that region (see Fig.~\ref{fig:TMur}),
implying that $S_\text{TP}$ is effectively a constant. Propagating the 
errors of $\tilde E^* = (-1.96 \pm  0.20)\times 10^{-3}$ and $\beta^*  = (-9.09 \pm 0.03 )\times 10^{-3}$ according to
\begin{equation}
\label{appeq:sigstp}
\sigma_S = \frac{1}{ |\beta | }    \sqrt{  \sigma_{\tilde E}^2 +  S_\text{TP}^2 \sigma_{\beta}^2 },
\end{equation} 
we obtain an estimate of $S^*_\text{TP}  = (0.216 \pm 0.022)$ for our model in the temperature and density regions shown in Fig.~\ref{fig:trho1d}.
\begin{figure}
 \centering
 \includegraphics[width=7.5cm]{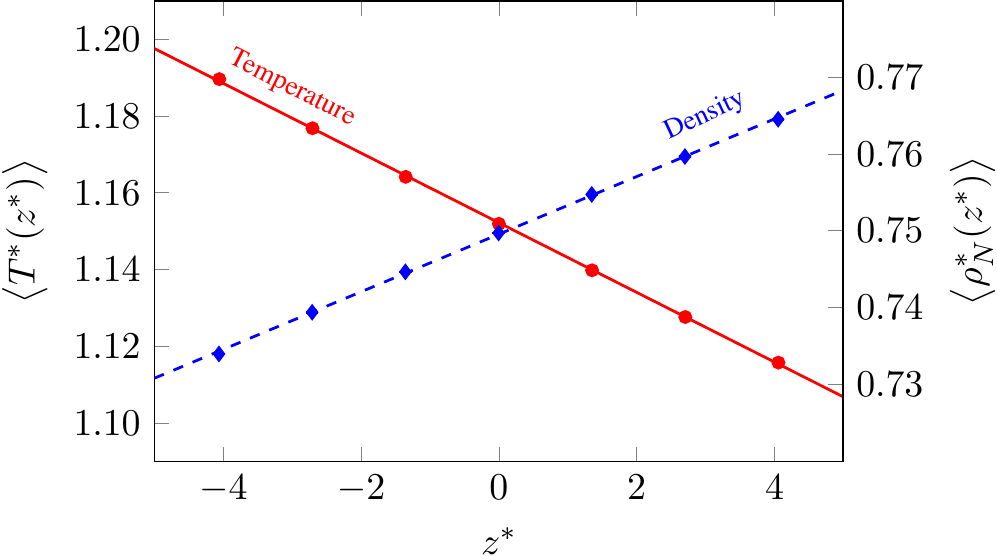}
  \caption{Temperature (red circles) and solvent number density (blue diamonds) averaged over slabs
           perpendicular to the $z$-axis in the vicinity of the origin. 
           The width of each slab is $\upDelta z^* = L^*/31$ and all error bars are smaller than the symbol sizes. 
           To estimate errors of the linear fit coefficients for the interval shown in the plot, 
           we first divided the NEMD trajectory into 1500 blocks and performed individual fits for each block average. 
           We then calculated the mean values and standard deviations of the resulting coefficients using block average analysis. 
           The results are $\big\langle T^*(z^*) \big\rangle = (-9.09 \pm 0.03 )\times 10^{-3} z^* + ( 1.1522 \pm 0.0002)$ (solid red line)
           and $ \big\langle \rho_N^*(z^*) \big \rangle =  (3.76  \pm 0.02) \times 10^{-3} z^* + ( 0.74952 \pm 0.00005)$ (dashed blue line).
         }
\label{fig:trho1d}
\end{figure}
\FloatBarrier

\section*{\textbf{References}}
\nocite{*}

%

\end{document}